\documentclass[preprint,12pt]{elsarticle}

\usepackage{amssymb}
\usepackage{amsmath}
\journal{Computer Methods and Programs in Biomedicine}

\begin{document}

\begin{frontmatter}

%% Title, authors and addresses

%% use the tnoteref command within \title for footnotes;
%% use the tnotetext command for theassociated footnote;
%% use the fnref command within \author or \affiliation for footnotes;
%% use the fntext command for theassociated footnote;
%% use the corref command within \author for corresponding author footnotes;
%% use the cortext command for theassociated footnote;
%% use the ead command for the email address,
%% and the form \ead[url] for the home page:
%% \title{Title\tnoteref{label1}}
%% \tnotetext[label1]{}
%% \author{Name\corref{cor1}\fnref{label2}}
%% \ead{email address}
%% \ead[url]{home page}
%% \fntext[label2]{}
%% \cortext[cor1]{}
%% \affiliation{organization={},
%%             addressline={},
%%             city={},
%%             postcode={},
%%             state={},
%%             country={}}
%% \fntext[label3]{}

\title{In silico modelling of changes in spinal cord blood flow after endovascular aortic aneurysm repair}

%% use optional labels to link authors explicitly to addresses:
%% \author[label1,label2]{}
%% \affiliation[label1]{organization={},
%%             addressline={},
%%             city={},
%%             postcode={},
%%             state={},
%%             country={}}
%%
%% \affiliation[label2]{organization={},
%%             addressline={},
%%             city={},
%%             postcode={},
%%             state={},
%%             country={}}

\author[label1,label6] {Michael Greshan Rasiah} %% Author name
\author[label2,label6]{Tom J A J Konings}
\author[label3]{Amanda Nio}
\author[label4]{Stefano Moriconi}
\author[label1]{Ashish S Patel}
\author[label1]{Alberto Smith}
\author[label1]{Panos Gkoutzios}
\author[label1]{Mohamed A Abdelhalim}
\author[label2]{Tammo Delhaas}
\author[label4]{M Jorge Cardoso}
\author[label3]{Pablo Lamata}
\author[label5]{Barend M E Mees}
\author[label1]{Bijan Modarai\corref{cor1}}
%% Author affiliation
\affiliation [label1]{organization={Academic Department of Vascular Surgery, St Thomas' Hospital, South Bank Section, School of Cardiovascular and Metabolic Medicine \& Sciences, King's British Heart Foundation Centre of Research Excellence, King's College London},%Department and Organization
            addressline={Westminster Bridge Road}, 
            city={London},
            postcode={SE1 7EH}, 
            country={United Kingdom}} 
\affiliation [label2]{organization={Department of Biomedical Engineering, Cardiovascular Research Institute Maastricht (CARIM), Maastricht University},%Department and Organization
            addressline={Universiteitssingel 50}, 
            city={Maastricht},
            postcode={6229 ER}, 
            country={the Netherlands}}
\affiliation [label3]{organization={Department of Digital Twins for Healthcare, School of Biomedical Engineering \& Imaging Sciences, St Thomas’ Hospital, King’s College London},%Department and Organization
            addressline={Westminster Bridge Road}, 
            city={London},
            postcode={SE1 7EH}, 
            country={United Kingdom}}
\affiliation [label4]{organization={Department of Biomedical Computing, School of Biomedical Engineering \& Imaging Sciences, King's College London},%Department and Organization
            addressline={Becket House, 1 Lambeth Palace Road, South Bank}, 
            city={London},
            postcode={SE1 7EU}, 
            country={United Kingdom}}
\affiliation [label5]{organization={Department of Vascular Surgery, Maastricht University Medical Centre+},%Department and Organization
            addressline={P. Debyelaan 25}, 
            city={Maastricht},
            postcode={6229 HX}, 
            country={the Netherlands }}
\cortext[cor1]{Corresponding author}
\ead{Bijan.Modarai@kcl.ac.uk}
\affiliation [label6]{country={Joint first authors}}

%% Abstract
\begin{abstract}
%% Text of abstract

Aims: To develop an in-silico model of the aorta and its spinal cord–supplying branches, and to characterise haemodynamic changes following aortic aneurysm (AA) repair with such a model. The work is motivated by the risk of spinal cord ischaemia (SCI) and paraplegia, serious complications that can arise from disruption of spinal cord perfusion during AA surgery. An objective, patient-specific tool capable of predicting changes in spinal cord blood flow prior to intervention would address a critical unmet clinical need. 

Methods: SimVascular was used to retrospectively create models of a 76-year-old female patient’s aorta pre- and post- uncomplicated endovascular AA repair. The full extent of the aorta and its branches, including vessels supplying the spinal cord, was segmented. Pulsatile flow simulations were conducted under the assumption of rigid vessel walls, with patient-specific inlet and three-element Windkessel models for the outlet boundary conditions on the SimVascular Gateway Cluster. Haemodynamic changes following stent graft implantation were evaluated, along with key surface-based metrics: time-averaged wall shear stress (TAWSS), oscillatory shear index (OSI), relative residence time (RRT) and endothelial cell activation potential (ECAP) were assessed with the primary focus on spinal cord–supplying vessels.  

Results: Postoperatively, segmental artery flow to the spinal cord decreased by 51.86\% due to exclusion of lumbar and posterior intercostal arteries by the stent graft. Spinal cord–supplying arteries showed increased TAWSS (+5.2\%) and reduced RRT and ECAP, with minimal change in OSI. Consistent with redistribution away from the spinal territory, modest postoperative flow increases were observed in non-spinal vascular beds, including the legs (+6.09\%), reno-visceral vessels (+5.89\%), and supra-aortic branches (+5.97\%). Across vascular territories, visceral arteries had the highest TAWSS and lowest RRT/ECAP, while leg arteries had the lowest TAWSS and highest RRT/ECAP; supra-aortic vessels exhibited the highest OSI. 

Conclusion: This study lays a foundation for computational prediction of SCI risk. It leverages in-silico modelling, using an open-source software pipeline and routine medical imaging, to assess spinal cord blood flow alterations after aortic surgery. Scaling to more patients and enriching the physiological detail of models may forge a path toward a clinical decision-making tool.

\end{abstract}

%%Graphical abstract
\begin{graphicalabstract}

\includegraphics [width=13cm]{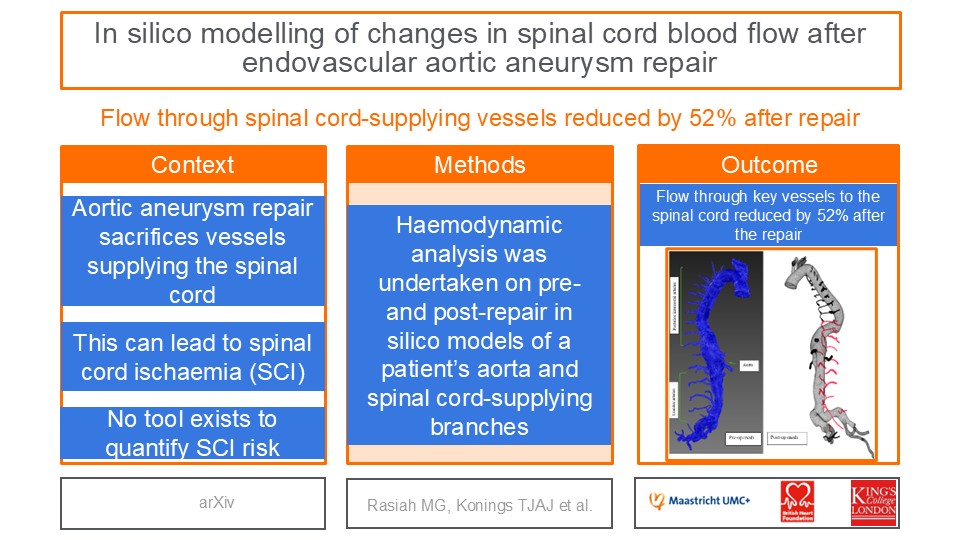}
\end{graphicalabstract}

%%Research highlights
\begin{highlights}
\item An aortic aneurysm model with spinal cord-supplying branches was created in silico.  
\item Flow through spinal cord-supplying vessels reduced by 52\% after simulated endovascular repair.
\item This model provides a foundation for a tool that objectively predicts changes in spinal cord blood flow prior to aortic aneurysm repair, addressing a significant unmet clinical need.
\end{highlights}

%% Keywords
\begin{keyword}
Spinal cord ischaemia, thoracoabdominal aortic aneurysm, aortic surgery, endovascular surgery.
%% keywords here, in the form: keyword \sep keyword

%% PACS codes here, in the form: \PACS code \sep code

%% MSC codes here, in the form: \MSC code \sep code
%% or \MSC[2008] code \sep code (2000 is the default)

\end{keyword}

\end{frontmatter}

%% Add \usepackage{lineno} before \begin{document} and uncomment 
%% following line to enable line numbers
%% \linenumbers

%% main text
%%

\section{Introduction}

An aortic aneurysm (AA) is an abnormal dilation of the aorta that carries a high risk of rupture. To prevent rupture, fatal in approximately two-thirds of cases, more than 90, 000 prophylactic repair procedures are performed annually worldwide \cite{howard2015population, castro2019disparities}. The mainstay of treatment for extensive aneurysms that involve both the abdominal and thoracic portions of the aorta, so-called thoracoabdominal aneurysms, is to reline the aneurysmal section with a stent graft, excluding the aneurysm sac from the circulation.

Paraplegia is an unpredictable complication of this procedure. This is becase whilst large calibre vessels supplying abdominal organs are preserved through fenestrations or branch grafts, some spinal cord-supplying vessels must inevitably be excluded from circulation. This exclusion can lead to spinal cord ischemia (SCI), and subsequent neurological deficits. Paraplegia is difficult to predict based on the extent of coverage and the number of vessels sacrificed alone \cite{verzini2014current, aucoin2023predictors, abdelhalim2022multicenter, rasiah2024need}.  Pre-operative planning of stent grafts typically includes analysis of computed tomography (CT) aortograms. Currently, however, this standard analysis cannot objectively predict paraplegia \cite{dias2015short}. Patients are counselled pre-operatively using anecdotal evidence which deters some from life-saving surgery. Up to 60\% of those who develop paraplegia after surgery remain wheelchair-bound, half will need a long-term urinary catheter, a third develop chronic pain and over one quarter of patients are dead after a year \cite{SpinalCordInjuryStatisticalCenter2021, rasiah2024optimizing, nana2025editor, rasiah2021medical}. 

In this context, a personalised tool that objectively determines the risk of paraplegia, accounting for the patient’s specific anatomy and the proposed repair strategy, is a significant unmet need. Central to predicting SCI post AA repair is the need to better understand blood flow to the spinal cord, and how exactly this is disrupted by introducing a stent graft. We address this need with digital twin technology, with the creation of a patient-specific model integrating anatomical (CT aortogram) and physiological data (cardiac output, heart rate), allowing computational fluid dynamics (CFD) simulation (with SimVascular) of the stent intervention and its consequences. Specifically, our main goal was to predict spinal cord blood flow after the stent placement.

Recent studies suggest that diffuse atherosclerosis affecting blood vessels supplying the spinal cord, such as the lumbar arteries, may contribute to SCI risk after aortic repair. Additionally, thromboembolic phenomena, potentially influenced by disturbed haemodynamics, could further compromise spinal cord perfusion \cite{aucoin2023predictors, lv2022artificial, carpenter2023nonlinear, kauppila2009atherosclerosis}. Quantifying these wall shear stress-based metrics may therefore reveal important mechanisms of SCI predisposition, offering a complementary layer of analysis uniquely enabled by CFD techniques. In this context, our secondary goal was to evaluate the impact of the surgery in surface-based haemodynamic metrics: time-averaged wall shear stress (TAWSS), oscillatory shear index (OSI), relative residence time (RRT), and endothelial cell activation potential (ECAP). These metrics provide insight into the mechanical forces acting on the vessel wall and have been implicated in both atherosclerotic plaque development and thromboembolic risk. TAWSS reflects the average tangential force exerted by blood flow on the endothelium and stent-graft, with both excessively high and low values linked to vascular pathology \cite{nichols2022mcdonald}. OSI quantifies flow reversal \cite{sotelo20163d} and RRT captures the duration blood particles remain near the vascular wall; conditions that may promote thrombus formation \cite{riccardello2018influence}. ECAP combines these factors to further assess thrombogenic potential \cite{mutlu2023does}.

In summary, we present the construction of a digital twin of spinal cord-supplying vessels based on CT aortograms from a patient undergoing endovascular thoracoabdominal AA repair. This model was built to predict the changes in blood flow to the spinal cord, and to infer key wall shear stress-derived metrics to assess the haemodynamic environment of the spinal cord collateral network and its potential role in SCI predisposition.

\section{Methods}

\subsection{Clinical and imaging data}

\begin{figure*}[h]
\centering
\includegraphics[width=12cm]{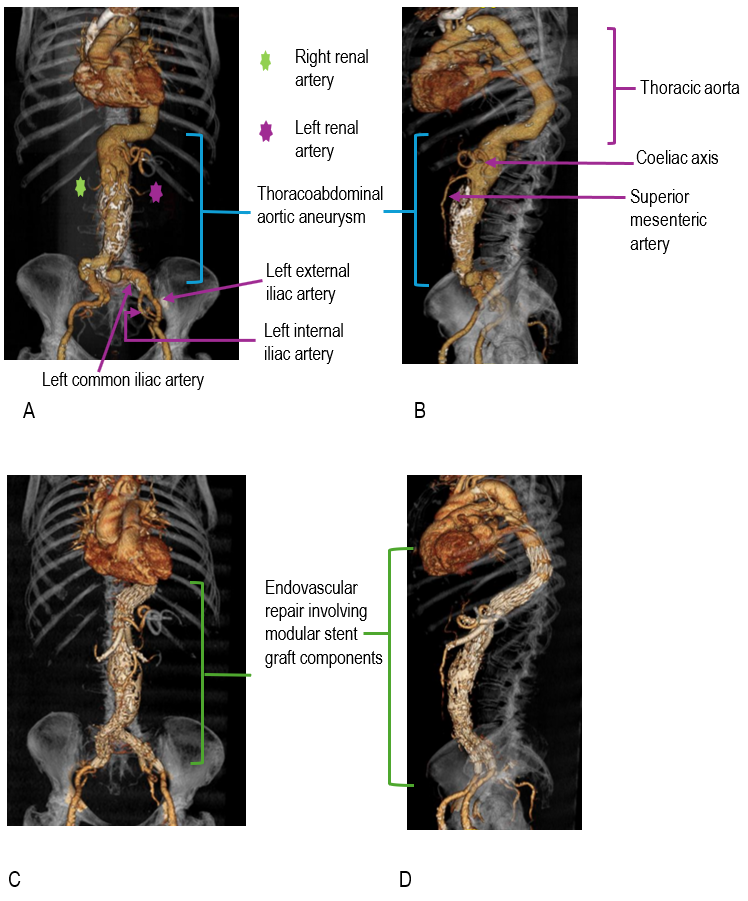}
\caption{The CT aortogram for the patient used to create the model in the present study (with organ structures not pertinent to spinal cord blood supply removed). \textbf{A}: Pre-operative anterior-posterior (AP) view. \textbf{B}: Pre-operative lateral view. \textbf{C}: Post-operative AP view with modular stent graft components in situ. Stent graft components include a tBranch device, a proximal thoracic stent, and separate branches to incorporate the coeliac, superior mesenteric and renal arteries. Bilateral iliac branch devices incorporate both internal iliac arteries. \textbf{D}: Post-operative lateral view with stent graft in situ. }
\label{Figure_1}
\end{figure*}

Analysis was undertaken using CT imaging from a 76-year-old female patient with a thoracoabdominal AA extending from the thoracic aorta at the level of the 8\textsuperscript{th }thoracic vertebra to both common iliac arteries \textbf{(Figure \ref{Figure_1} A, B)}. The endovascular repair entailed placing modular stent graft components from the level of the thoracic aorta to both external iliac arteries. A tBranch device (Cook Medical, Bloomington, IN, USA) \cite{tsilimparis2017technical} was used for the abdominal segment with separate branches used to incorporate the coeliac, superior mesenteric and renal vessels \textbf{(Figure \ref{Figure_1} C, D)}. Bilateral iliac branch devices (Cook Medical, Bloomington, IN, USA)  \cite{de2019use} were used to incorporate both internal iliac arteries. The procedure was technically successful, with all target vessels patent and no evidence of SCI.

Arterial phase contrast-enhanced (Omnipaque 300 x 100mls) CT aortograms were performed pre- and one month post-operatively from the root of the neck to the greater trochanter. There were 675 axial images of 1mm slice thickness in the pre-op scan. The tube current varied between 290mA and 440mA, whilst the tube voltage was 120kV. Images were stored as Digital Imaging and Communications in Medicine (DICOM) files on a Picture Archiving and Communication System (PACS) server.

 \subsection{Software, operating system and cluster}

SimVascular, an open-source software, was used as a comprehensive pipeline for imaging-derived segmentation of pertinent vasculature and blood flow simulation \cite{updegrove2017simvascular}. SimVascular was run on a Windows operating system. Owing to the complexity of the model created, a high-performance open-source computer cluster specifically created for SimVascular known as the Gateway was used to run simulations \cite{wilson2018using}. Paraview was used to visualise simulation results and undertake further post-processing to derive surface metrics of interest \cite{ahrens200536, ayachit2015paraview}.

\subsection{Image segmentation and geometry construction}

\begin{figure*}[h]
\centering
\includegraphics[width=13cm]{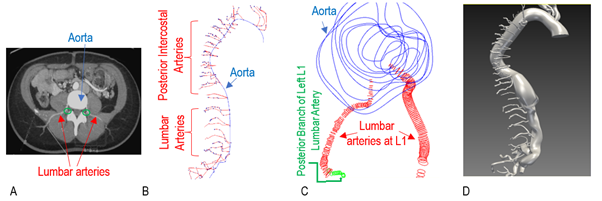}
\caption{Schematic demonstrating pipeline from image upload to solid model creation of the aorta and pertinent vessels (e.g. posterior intercostal arteries, lumbar arteries etc.) supplying blood to spinal cord. \textbf{A}: The pipeline begins by uploading medical imaging data in DICOM format to SimVascular. In this axial slice \cite{stillaert2023lumbar}, the aorta and two lumbar arteries, the posterior branches (in green circles) of which will supply the spinal cord, are shown. \textbf{B}: Individual points are plotted along vessels which feed the spinal cord to create centrelines or ‘Paths’ in SimVascular. Here the longitudinal blue path is the centreline for the aorta, and the approximately perpendicular red branches are of selected daughter vessels which supply the spinal cord. \textbf{C}: The lumen of each of the aforementioned ‘Paths’ are drawn out to create contour groups as part of the segmentation process in SimVascular. Here the contours of a section of the aorta (the largest blue contours) and the lumbar arteries at the level of L1 (red), as well as the posterior branch of the left L1 lumbar artery (green), are shown. This process is replicated for all vessels supplying the spinal cord. \textbf{D}: After lofting of segmentations, a model of the aorta and spinal cord-supplying vessels is created (capped appropriately to allow setting of boundary conditions), on which meshing and flow simulations can be undertaken.}
\label{Figure_2}
\end{figure*}

Vascular segmentation was performed in SimVascular using a contour-based approach. Centrelines (paths) were manually defined for the aorta, its major branches, and spinal cord-supplying vessels based on axial, sagittal, and coronal views, in which crosshairs were positioned at the vessel centres. SimVascular connected the path points in the investigator-defined order, generating a named path for each vessel. Contours were manually placed along each vessel to define lumen boundaries, and grouped into contour sets for subsequent model construction. \textbf{Figure \ref{Figure_2} A-C} schematically demonstrates the pipeline to segmentation. Thrombosed segments, identified by the absence of contrast enhancement, were excluded from the final model.

Solid models were generated in SimVascular by combining contour groups into lofted surface representations of the vessels of interest \textbf{(Figure \ref{Figure_2} D)}. Planar surfaces (caps) were added at vessel openings to produce water-tight geometries. The final model consisted of two face types: walls, representing vessel surfaces, and caps, which defined inlet and outlet boundaries. The PolyData modeller was used to generate triangle-based surface meshes. Although SimVascular also supports OpenCASCADE for NURBS-based modelling, PolyData was chosen due to its wider adoption and compatibility with the intended simulations \cite{updegrove2017simvascular}.

\subsection{Mesh construction}

The solid model was discretised into a tetrahedral finite element mesh using SimVascular. A global maximum edge size of 0.2043mm was used as a baseline, with local adjustments made to accommodate the wide variation in vessel calibres, particularly the small branches arising from posterior intercostal arteries.

\begin{figure*}[h]
\centering
\includegraphics[width=10cm]{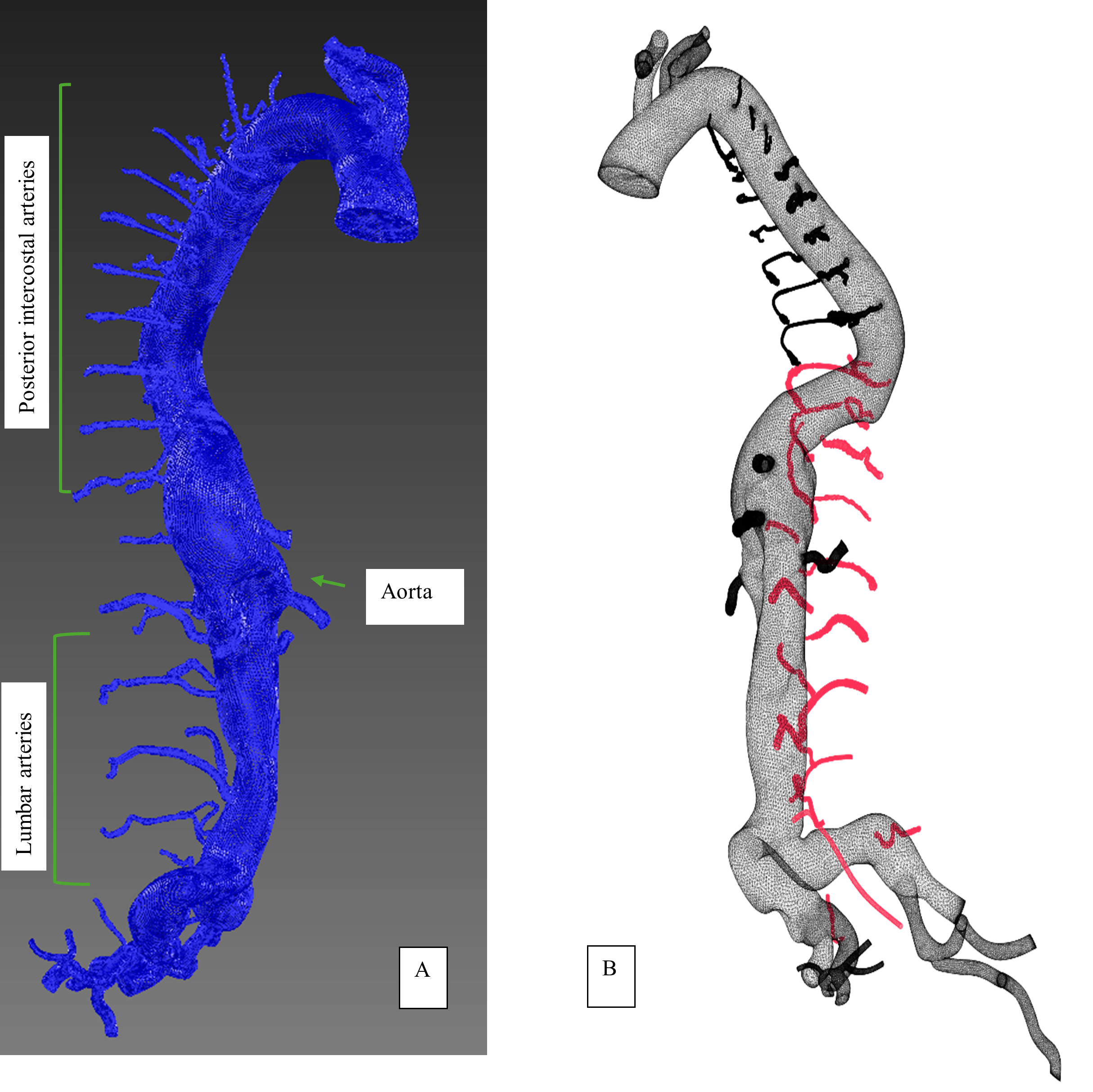}
\caption{\textbf{A:} Meshed model of aneurysmal aorta (viewed from front) with branches supplying the spinal cord created via SimVascular. \textbf{B:} The impact of stent insertion on the model with respect to vessel coverage. In red are the vessels which the post-operative scan showed were covered after the insertion of the stent device. Rather than creating a new model from scratch using the post-operative scan (which is not something that would be possible in a future pipeline that seeks to predict the risk of spinal cord ischaemia in patients about to undergo surgery), the pre-operative model was altered to reflect post-operative changes (something which a future predictive pipeline could indeed incorporate, since information generated when designing stents enables calculations of which branches of the aortic tree are likely to be covered).}
\label{Figure_3}
\end{figure*}

To address challenges posed by complex vascular geometry, a staged meshing strategy was employed, starting with smaller subregions and progressively extending to the full model. Vessel-specific edge lengths were refined iteratively to ensure mesh continuity and quality across the model. The final mesh consisted of approximately 4.34 million nodes and 25.49 million elements \textbf{(Figure \ref{Figure_3} A)}.

\subsection{Post-operative model}

We model the impact of the stent by a total occlusion of branches covered by the stent. This approach effectively represents the impact of the surgery without explicitly modelling the stent structure \textbf{(Figure \ref{Figure_3} B)}. As such, the number of totally occluded branches is the only input needed to perform simulation of the post-operative scenario. This strategy was adopted since any future pipeline for predicting SCI prior to surgery will only be reliant on the pre-operative CT and the planned extent and location of the stent. In our study, the post-operative occlusion of downstream branches was measured from the CT aortogram carried out after stent graft placement \textbf{(Figure \ref{Figure_1} C, D)}. 

\subsection{Inlet boundary conditions}

An aortic flow waveform was adapted from published data obtained from a female patient of comparable age and cardiovascular status \cite{zhao2023aortic} for pre- and post-operative simulations. The profile was scaled to match the cardiac output and heart rate of the patient in the present study. The waveform was applied at the inlet as a time-varying Dirichlet boundary condition, using a parabolic velocity profile over one cardiac cycle (1.11 s).

\subsection{Outlet boundary conditions}

Three-element Windkessel models (RCR: resistance-compliance-resistance) were implemented at each outlet \cite{westerhof1969analog}. 

A key aspect of the present study was the simulation of realistic flow rates to the vessels supplying the spinal cord, specifically the posterior intercostal and lumbar branches. However, in-vivo flow rates to these branches remain largely unknown. To estimate these flows, we relied on a combination of literature-derived clinical data and informed estimations, the details of which are outlined below. Since flow distributions for most other branches are available, albeit sparse, we first determined their flow rates using existing clinical data from the literature. The spinal cord-supplying branches (i.e. posterior intercostal arteries, lumbar arteries etc.) could then be derived by subtracting the known branch flows from the total aortic flow.

A comprehensive analysis of flow distribution to vessels visible on CT was performed based on physiological data and validated against multiple literature sources. In cases where no literature data were available for specific small branches, Murray’s law was employed to estimate flow distribution based on vessel diameters. At such bifurcations, Murray’s law assumes that flow $(Q)$ is proportional to the cube of the vessel radius $(r)$:

 \begin{equation}
    Q \propto  r^3,
    \label{Murray}
\end{equation}

which allowed us to assign relative flow rates in proportion to measured vessel diameters in the absence of empirical data.

%For a bifurcation where a parent vessel splits into two daughter vessels, the relationship is:  

%\begin{equation}
 %  (r_p)^3 = (r_1)^3 + (r_2)^3,
%    \label{Murray_2}
%\end{equation}
%where $r_p$ is the radius of the parent vessel and $r_1$ and $r_2$ are the radii of the daughter vessels.

As shown in   \textbf{Table \ref{Table_1}}, the total cardiac output was assumed to be 5L/min (83.33cm³/s), with distribution carefully allocated among major vessel groups.

\setlength{\tabcolsep}{4.5pt}
\begin{table}[htbp] 
\caption{\label{tab:font} Windkessel model parameters and vessel flow allocation. Proximal resistance $(R_1)$, distal resistance $(R_2)$ and compliance $(C)$ parameters as well as calculated flow allocations (as a percentage of cardiac output) are listed for selected vessels included in the model in the supra-aortic, visceral, renal, lower extremity, and spinal cord-supplying arteries.}
\vspace{4 mm}
\centerline{\begin{tabular}{lcccr} \hline\hline
\textbf{Vascular Territory /}& \textbf{$R_1$}& \textbf{$R_2$}&\textbf{$C$}& \textbf{Flow}\\
 Branch& & & &\textbf{(\%)}\\ \hline
\textbf{Supra-aortic Vessels}& & && \\
Brachiocephalic Trunk& 0.11& 0.96&1.68
& 15\\
Left Common Carotid& 0.21& 1.92&0.84
& 7.5\\
Left Subclavian& 0.21& 1.92&0.84
& 7.5\\
\textbf{Visceral Arteries}& & &
& \\
Coeliac Trunk& 0.16& 1.40&1.15
& 10.25
\\
Superior Mesenteric Artery (SMA)& 0.12& 1.08&1.49
& 13.25
\\
\textbf{Renal Circulation}& & &
& \\
 Renal Left& 0.17& 1.51& 1.07
&9.5
\\
 Renal Right& 0.17& 1.51& 1.07
&9.5
\\
\textbf{Lower Extremity – Left}& & & 
&\\
 Left External Iliac& 0.23& 2.05& 0.79
&7\\
 Left Superior Gluteal& 3.19& 28.73& 0.06
&0.5\\
 Left Inferior Gluteal & 1.60& 14.36& 0.11
&1\\
\textbf{Lower Extremity – Right}& & & 
&\\
 Right External Iliac& 0.23& 2.05& 0.79
&7\\
 Anterior Trunk of Right Internal Iliac& 1.60& 14.36& 0.11
&1\\
 Spinal Branch of Right Iliolumbar& 63.84& 574.56& 0.00
&0.025\\
 Posterior Trunk of Right Internal Iliac& 3.36& 30.24& 0.05
&0.475\\
 \textbf{Thoracoabdominal Spinal Cord-supplying Arteries}& & & &\\
 Posterior Intercostal (each)& 0.03& 5.91& 53.15&0.28\\
 Lumbar (each)& 0.03& 5.91& 53.15&0.28\\
\end{tabular}}
\label{Table_1}
\end{table}

\subsubsection{Supra-aortic vessels}

The brachiocephalic trunk was allocated 15.00\% of total flow, which aligns with clinical findings indicating 15-20\% of cardiac output directed to carotid arteries \cite{xing2017distribution, benim2011simulation}. The left common carotid and left subclavian arteries each receive 7.50\% of flow, supported by clinical data showing approximately 8\% flow to subclavian arteries \cite{lantz1981regional}.

\subsubsection{Visceral arteries}

The coeliac trunk receives 10.25\% of total flow \cite{shiozawa2022blood}, comprised of 3\% of cardiac output to the stomach, 1\% to the spleen, and 6.25\% to the hepatic artery. 

The superior mesenteric artery (SMA) receives 13.25\% \cite{blanco2014blood} of total flow, consisting of 3.25\% to large intestine and 10\% to small intestine. Some clinical studies report 15.4\% flow \cite{lantz1981regional}, with variations due to prandial conditions.

\subsubsection{Renal circulation}

The renal arteries each receive 9.50\% of total flow. This aligns with clinical data from the literature showing 8.6-10\% flow per kidney \cite{blanco2014blood}, with a total renal perfusion of 1-1.2L/min for both kidneys.   

\subsubsection{Lower extremity circulation}

Each leg receives 8.50\% of the total flow  \cite{blanco2014blood}, with the external iliac artery supplying most of the lower limb and the internal iliac artery perfusing the pelvis and gluteal region. In the left leg, 7\% flows through the left common and external iliac arteries \cite{itzchak1975external}, while 1.5\% is distributed via the left internal iliac artery \cite{bjorck2006blood}. The right leg follows the same pattern. 

%Lumbar circulation is integrated by subtracting iliolumbar artery flow from the anterior trunk, ensuring accurate perfusion modelling. 

%Murray’s law, which states that flow $(Q)$ in a vessel is proportional to the cube of its radius $(r)$:  
%\begin{equation}
%    Q \propto  r^3,
 %   \label{Murray}
%\end{equation}
%was applied at bifurcations to maintain physiologically realistic flow based on vessel diameters.

%For a bifurcation where a parent vessel splits into two daughter vessels, the relationship is:  
%\begin{equation}
%   (r_p)^3 = (r_1)^3 + (r_2)^3,
%    \label{Murray_2}
%\end{equation}
%where $r_p$ is the radius of the parent vessel and $r_1$ and $r_2$ are the radii of the daughter vessels.

\subsubsection{Spinal vessel flow distribution}

After accounting for all major vascular territories described above (supra-aortic: 30\%, visceral: 23.5\%, renal: 19\%, and leg arteries: 17\%), approximately 10\% of the total cardiac output remains for the spinal cord-supplying vessels, predominantly consisting here of the posterior intercostals and the lumbars. This distribution aligns with physiological expectations and can be verified through the flow distribution calculations shown in \textbf{Table \ref{Table_1}}. In the present model, an approach of equal flow distribution among the spinal vessels is adopted, recognising that whilst there may be minor variations in individual vessel perfusion, the total flow to the spinal cord must be maintained within physiological limits. This assumption of equal distribution is implemented through the parallel arrangement of identical RCR elements at each spinal branch outlet, such that the overall resistance of the territory matched the target ($R_{spinal}$)

\subsubsection{RCR parameter calculation}

The terminal resistance $(R_t)$ for each branch was calculated from the ratio of mean arterial pressure to the prescribed branch flow:   

\begin{equation}
    R_t=\frac{\bar{P}}{Q_i} ,
     \label{Terminal_Resistance}
\end{equation}
where $\bar{P}$ represents the mean arterial pressure (13.3kPa) and $Q_i$ is the flow at the branch $\textit{i}$.   

The two resistive components of the three-element Windkessel model were defined as fractions of $R_t$:

\begin{equation}
    R_1=0.1R_t, R_2=0.9R_t, 
     \label{Prox_Distal_Resistance}
\end{equation}
where $R_1$ and $R_2$ are proximal and distal resistances respectively, summing to $R_t$.

%such that $R_1+R_2=R_t$. The compliance $(C)$ was calculated using:

%\begin{equation}
 %   R_1=0.1(R_1+R_2)=0.9R_t, 
 %    \label{Prox_Distal_Resistance}
%\end{equation}
The compliance was determined from the diastolic decay time constant, $\tau$, as:

\begin{equation}
C=\frac{\tau}{R_t},
     \label{Compliance}
\end{equation}
where $\tau$ = 1.79 s \cite{xiao2014systematic}.

The full set of outlet parameters ($R_1$, $R_2$, and $C$) are summarised in \textbf{Table \ref{Table_1}} for  vessels included in this model. These values yield physiologically plausible haemodynamic responses while maintaining numerical stability.

To isolate the effects of anatomical variation, identical RCR parameters were applied to both pre- and post-operative models, without introducing any adaptive changes in the downstream vasculature.

\subsection{Simulation and postprocessing}

The simulation employed standard blood properties in CGS units; dynamic viscosity was 0.04 poise (gr$\cdot$cm$^{-1}$$\cdot$s$^{-1}$) and blood density was 1.06 (gr$\cdot$cm$^{-3}$).

Pulsatile flow simulations on the pre- and post-operative models were run with an incompressible Navier-Stokes flow solver. Walls were set as rigid. In order to achieve periodicity, five cardiac cycles were first simulated (cycle by cycle) at low temporal resolution, before a 6\textsuperscript{th }cardiac cycle was simulated at high resolution, with the last step from each simulation saved and used as the restart file for the next simulation. Following literature published by developers of SimVascular \cite{tran2024patient}, the time step size was set to 1/500\textsuperscript{th} of a cardiac cycle. The residual criteria was set to 1 x 10\textsuperscript{-4}. Running a simulation on a large mesh, in this case with over 20 million elements, is computationally demanding and, therefore, requires access to a high-performance computing (HPC) cluster. Gateway \cite{updegrove2017simvascular}, SimVascular’s own HPC cluster, was utilised to run simulations on the model. The time taken for simulations was between 4 (post-operative model) and 13 (pre-operative model) hours, as the post-operative model contains fewer elements after vessel occlusion. ParaView was then used to visualise the results and derive the surface metrics of time-averaged wall shear stress (TAWSS), oscillatory shear index (OSI), relative residence time (RRT) and endothelial cell activation potential (ECAP).
 
\subsection{Ethical approvals}

Ethics was secured for this work at King’s Health Partners (KHP) (REC: 13/LO/1685).

\section{Results}

\subsection{Haemodynamic changes after stent graft implantation}

\begin{figure*}[h]
\centering
\includegraphics[width=13cm]{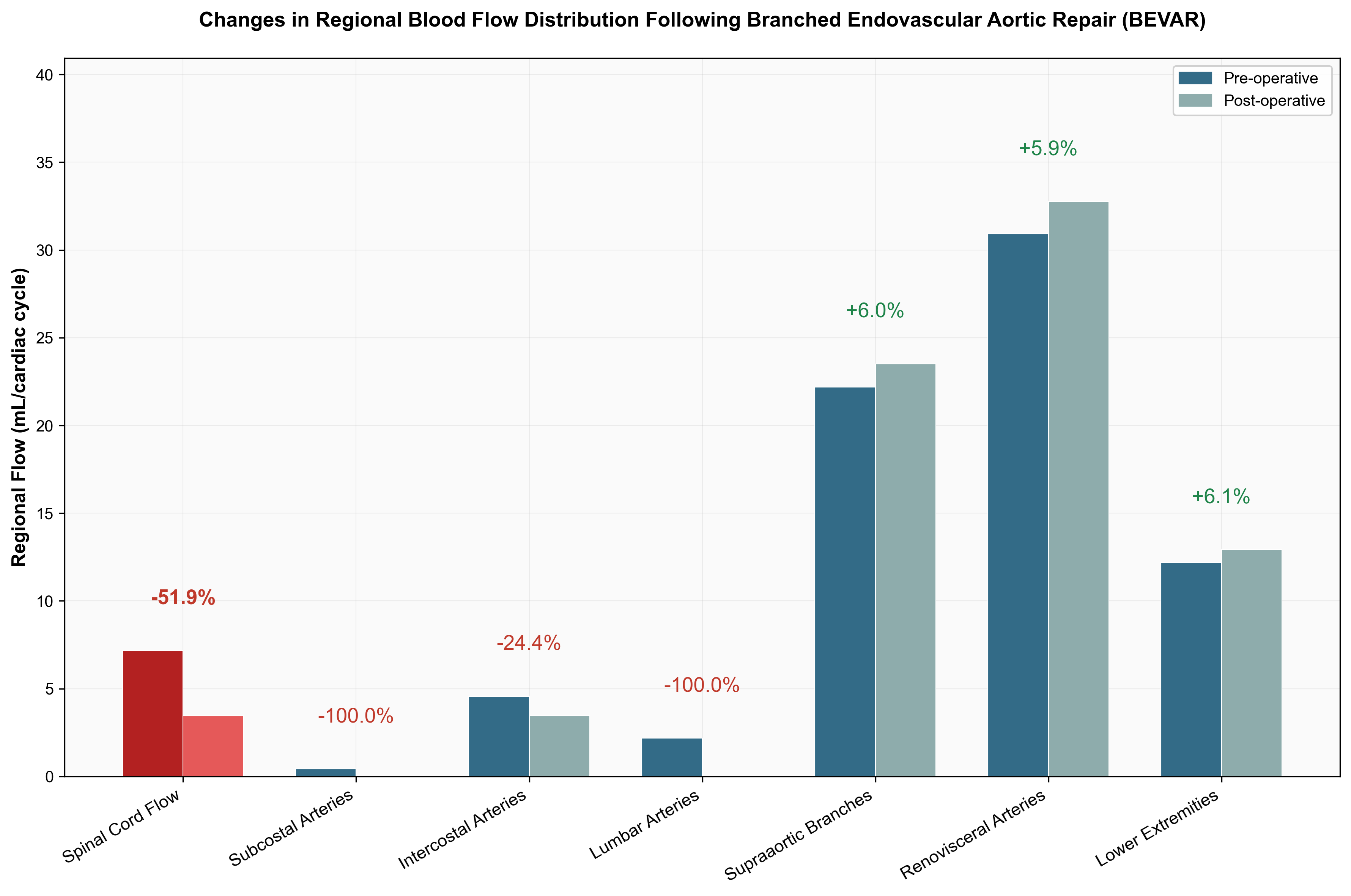}
\caption{Regional blood flow distribution (mL/cardiac cycle) before and after branched endovascular aortic repair (BEVAR). Bar pairs represent pre-operative (dark red/blue) and post-operative (light red/blue) flow in key vascular areas, including spinal cord, subcostal, intercostal, lumbar, supra-aortic, renovisceral, and lower extremity regions. Percent changes are annotated above each region. Notably, complete flow loss (-100\%) occurred in the subcostal and lumbar arteries, and spinal cord flow decreased by 51.9\%. In contrast, flow increased modestly in supra-aortic (+6.0\%), renovisceral (+5.9\%), and lower extremity (+6.1\%) regions postoperatively.}
\label{Figure_4}
\end{figure*}

The posterior intercostal arteries and lumbar arteries serve as the source for many of the segmental medullary and radicular arteries that supply the spinal cord. As imposed by the condition of the post-operative model, the lumbar arteries were all excluded after stent graft implantation, with 100\% reduction in flow post-operatively \textbf{(Figure \ref{Figure_4})}. This occlusion following the placement of a stent graft caused the total spinal cord perfusion to decrease from 7.18 ml/cycle to 3.46 ml/cycle (a reduction of 51.86\%). Complementarily, and as expected by the conservation of flow, other vascular territories experienced modest increases in perfusion to compensate this reduction of 3.72 ml/cycle: blood flow to the legs increased from 12.19 ml/cycle to 12.93 ml/cycle (an increase of 6.09\%); in the renovisceral vessels from 30.93 ml/cycle to 32.75 ml/cycle (an increase of 5.89\%); and in supra-aortic vessels from 22.18 ml/cycle to 23.51 ml/cycle (an increase of 5.97\%).

\subsection{Surface metrics}

\begin{figure*}[h]
\centering
\includegraphics[width=7cm]{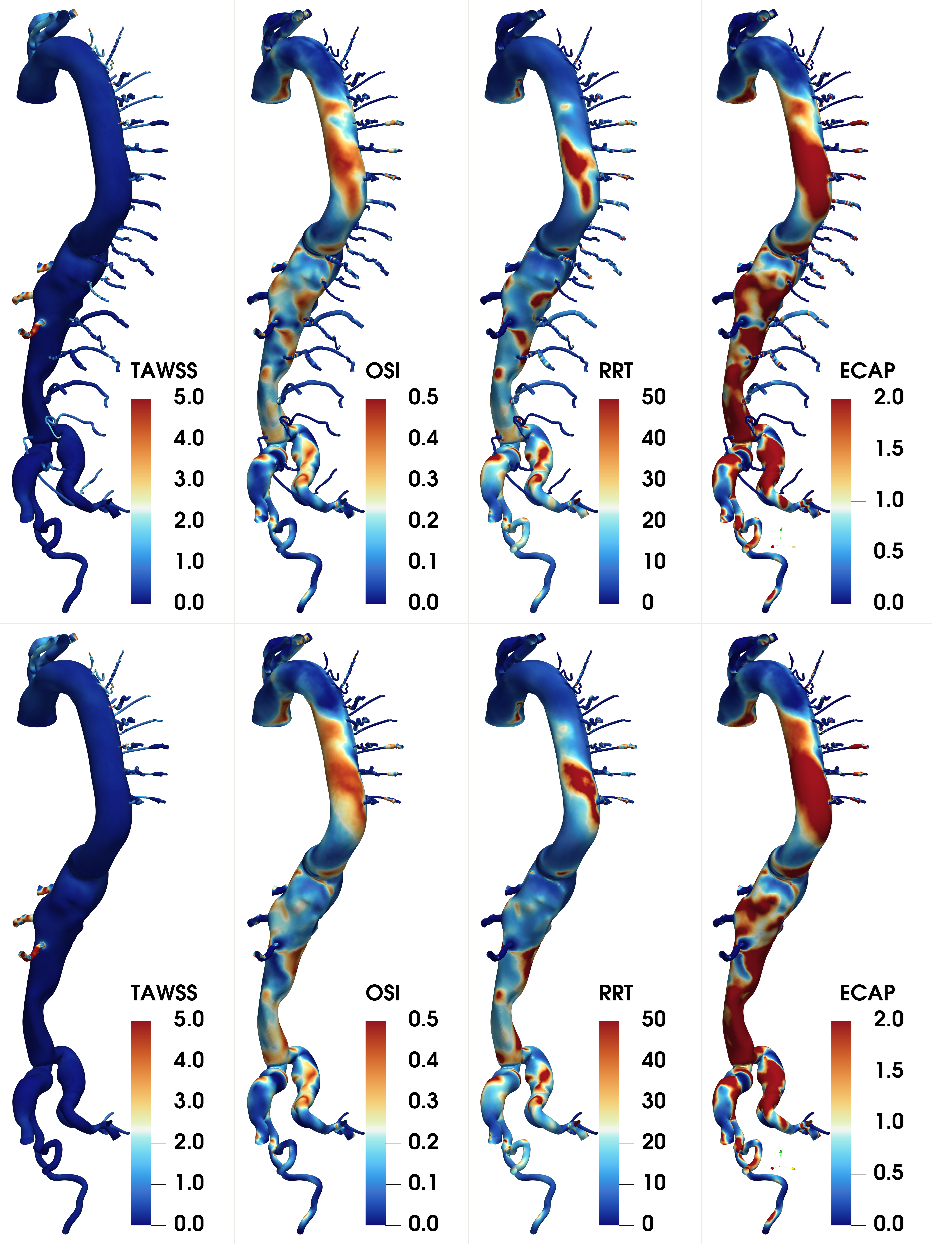}
\caption{Surface metrics visualisation. The figure presents three surface metrics; TAWSS (1st column), OSI (2nd column), RRT (3rd column) and ECAP (4th column) are shown for pre- and postoperative states (top and bottom rows, respectively). The vascular walls are colour coded according to the respective parameter values, as indicated by the accompanying colour bars. These haemodynamic metrics provide insight into regions susceptible to thromboembolic events and vascular disease.}
\label{Figure_5}
\end{figure*}

Wall shear stress-related haemodynamic parameters were evaluated in both pre-operative and post-operative simulations. The primary metrics included TAWSS, OSI, RRT, and ECAP. \textbf{Figure \ref{Figure_5}} presents a comprehensive visualisation of haemodynamic parameters in the spinal cord-supplying branches, comparing pre- and post-operative conditions, with branches colour coded according to parameter values. 

The postoperative model reflects the anatomical changes following stent-graft placement, which includes the exclusion of several segmental medullary and radicular arteries present in the preoperative state.

Across all patent vessels in both the pre-operative and post-operative models, the haemodynamic parameters occupied a wide range of values. TAWSS values ranged from 0.09 Pa to 4.74 Pa. OSI varied from 0.01 to 0.17, while RRT spanned from 0.43 s to 57.56 s. The calculated ECAP values ranged from 0.01 to 1.75. These global ranges reflect the diverse fluid dynamic environments throughout the aorta, from the high-flow visceral branches to the low-flow distal vessels of the lower extremities.

A primary focus of the analysis was the haemodynamic environment within the posterior intercostal and lumbar arteries supplying the spinal cord. Pre-operatively, these vessels exhibited TAWSS values between 0.12 Pa and 1.74 Pa and OSI values from 0.01 to 0.15. Following the intervention, the remaining patent posterior intercostal arteries showed a general trend of increased TAWSS and decreased RRT and ECAP. For instance, the left T3 posterior intercostal artery (LT3PI) experienced a 3.9\% increase in TAWSS (from 1.74 Pa to 1.81 Pa), accompanied by a 5.5\% decrease in RRT. Similarly, the right T4 posterior intercostal artery (RT4PI) showed a 4.9\% increase in TAWSS. The surface-averaged TAWSS for the entire spinal territory increased by 5.2\%, from 0.81 Pa to 0.85 Pa postoperatively. Changes in OSI were minimal across the patent segmental medullary-feeder arteries.

When haemodynamic parameters were surface-averaged across distinct vascular territories, clear regional patterns emerged. Pre-operatively, the visceral arteries were characterised by the highest mean TAWSS (3.42 Pa), whereas the leg arteries exhibited the lowest (0.12 Pa). Conversely, RRT and ECAP were highest in the leg arteries (19.40 s and 1.06, respectively) and lowest in the visceral territory (0.96 s and 0.07). The supra-aortic branches presented the highest mean OSI (0.09). Post-operatively, mean TAWSS increased in the visceral (8.1\%), supra-aortic (2.0\%), and spinal (5.2\%) territories, while remaining nearly constant in the leg arteries. RRT and ECAP decreased across all territories, with the most substantial relative reduction observed in the spinal region.
   
\section{Discussion}

This is one of the first studies to computationally model spinal cord-directed blood flow in a patient before and after endovascular stent graft repair of a thoracoabdominal AA. We objectively quantified the haemodynamic consequences of sacrificing spinal cord-supplying vessels and observed a reduction in spinal perfusion of approximately 50 percent following stent deployment in this patient-specific case. This personalised, image-based CFD pipeline provides a foundational platform for understanding the haemodynamic basis of SCI. It also supports the potential development of a predictive risk stratification framework that incorporates haemodynamic rather than purely anatomical criteria.

\subsection{Simulations to enrich anatomical scans}

A simulation platform that predicts SCI risk represents a meaningful step forward compared to existing clinical strategies, which predominantly rely on anatomical scoring systems that estimate risk based on the number of excluded segmental medullary and radicular arteries \cite{chung2022prevention}. These methods do not consider the actual functional contribution of those vessels to spinal cord blood flow and may therefore misrepresent individual risk. By integrating detailed patient-specific flow modelling, the proposed approach offers a more physiologically grounded method for evaluating the consequences of endovascular repair.

 All software used in the pipeline was open-source \cite{updegrove2017simvascular, wilson2018using, ahrens200536, ayachit2015paraview, fedorov20123d}, and the only data required was from routine pre-operative imaging. This makes the method inherently scalable for future clinical implementation, without the need for additional diagnostic procedures or associated costs. The focus on widely available inputs reinforces the translational relevance of this work and supports its feasibility for integration into standard care pathways.

 \subsection{Flow changes following stent grafting}

Following stent graft deployment, spinal cord perfusion decreased by 51.86\% (from 7.18 to 3.46 ml) due to segmental medullary and radicular artery exclusion. 

This is a significant step forward in understanding and predicting SCI post aortic surgery, as herein lies a quantitative means of appreciating blood flow to the spinal cord.

Nevertheless, the authors do not wish such a figure to be extrapolated prematurely. No in-vivo information exists regarding spinal cord perfusion to validate this claim (though to gain such validation would be extremely invasive, which is why in silico modelling was adopted to begin with). Additionally, whilst considerable thought was afforded to boundary conditions as described in the Methods section, changes in boundary conditions would change the output of flow simulations. Furthermore, to better appreciate the influence of collateral flow within the network of vessels that supplies the spinal cord, a (e.g. one- or zero-dimensional) model would need to be developed to account for such collateral vessels which are often not visible on CT aortograms.
 
The spinal cord perfusion reduction following stent graft deployment demonstrates the direct haemodynamic consequences of segmental medullary and radicular artery exclusion. This flow redistribution arises mechanically from the loss of outflow pathways, forcing blood toward remaining patent vessels. While three-dimensional geometric factors such as vessel curvature and diameter transitions also contribute to flow resistance, the dominant effect in this model is the elimination of outlet pathways. Without autoregulatory responses or collateral pathways from paraspinous and pelvic circulations, the model reveals hypothetical haemodynamic consequences of segmental medullary and radicular artery coverage based on the assumption of equal boundary conditions not disturbed by surgery, providing a quantitative appreciation of blood flow reduction in the absence of compensatory mechanisms.

 \subsection{Wall shear environment in spinal cord-supplying arteries}

The observed changes in spinal vessel wall shear stress metrics following stent grafting provide important insights into potential mechanisms of SCI predisposition, as hypothesised in our study rationale. Following stent deployment, spinal territory demonstrated increased TAWSS (+5.2\%) with concurrent decreases in OSI, RRT and ECAP, indicating reduced flow stagnation and decreased tendency for flow reversal in the remaining patent spinal vessels.

These findings are particularly relevant to our hypothesis that diffuse atherosclerosis affecting spinal cord collateral circulation may contribute to SCI risk. The increased TAWSS in remaining spinal vessels suggests compensatory flow redistribution that may place additional mechanical stress on collateral pathways, including the lumbar arteries identified as critical components of spinal cord blood supply. While elevated TAWSS can promote beneficial endothelial adaptation, it may also accelerate atherosclerotic progression in vessels with pre-existing disease, potentially compromising the spinal cord's collateral reserve.

The reduction in RRT and ECAP in patent spinal cord-supplying vessels indicates improved flow characteristics with reduced residence time near vessel walls. However, this improvement occurs only in vessels that remain patent post-stent grafting. In the context of our hypothesis regarding thromboembolic phenomena, it can be hypothesised that closure of segmental arteries may introduce thromboembolic risk through altered haemodynamics that promote stasis in collateral pathways or at arterial branch points. While our surface-based metrics suggest improved flow conditions in remaining vessels, the elimination of multiple segmental medullary and radicular arteries may create haemodynamic disturbances in areas not captured by our current modelling approach.

The clinical significance of these shear stress metrics extends beyond the immediate post-operative period. TAWSS and OSI are well-established indicators of vascular pathology, with TAWSS closely associated with aortic remodelling \cite{lv2022artificial} and OSI implicated in atherosclerosis development \cite{carpenter2023nonlinear}. Although the relationship between peripheral vascular disease and SCI following endovascular repair is still being elucidated, emerging evidence suggests that diffuse atherosclerotic burden, particularly within collateral vascular beds, may reduce the spinal cord's compensatory capacity \cite{aucoin2023predictors}. The documented association between lumbar artery atherosclerosis and degenerative spinal conditions \cite{kauppila2009atherosclerosis, beckworth2018atherosclerotic} may reflect more widespread impairment in spinal vascular supply. The patient-specific quantification of wall shear stress patterns enabled by this pipeline could therefore provide valuable data for evaluating how pre-existing vascular pathologies might influence SCI risk in individual patients.

\subsection{Regional haemodynamic phenotypes}

Surface-based flow analysis revealed distinct haemodynamic phenotypes across different vascular territories, with visceral regions showing the largest relative TAWSS increase (+8.1\%) compared to spinal territory (+5.2\%) and other regions. This differential response reflects the unique physiological characteristics and flow demands of each vascular bed.

Regionally, leg arteries maintained the lowest TAWSS and highest RRT and ECAP values, owing to low flow under simulated resting/non-exercise conditions. This reflects the well-described distribution of resting blood flow, where visceral organs receive a disproportionately higher share of perfusion compared to the limbs. Visceral vessels displayed the highest TAWSS and lowest RRT and ECAP, driven by a high-flow narrow-calibre configuration that raises velocity and shear stress. Whole-aorta CFD exposed these territory-specific shear phenotypes, underscoring the value of system-level modelling.

The visceral branch metrics suggest a distinct haemodynamic environment relevant to arterial remodelling and long-term stent graft patency post-intervention. In thoracoabdominal AA repair, visceral vessels are often revascularised using bridging stents or fenestrations, and the long-term success of these interventions depends partly on maintained flow conditions. Mechano-transductive forces characterised by TAWSS and ECAP influence endothelial behaviour and wall adaptation after thoracoabdominal AA repair. Whether these conditions promote beneficial remodelling or pathological change remains uncertain, but the observed differences justify including these territories in future analyses given the clinical significance of mesenteric ischaemia after aortic surgery.

\subsection{Clinical relevance and translational potential}  

The integration of patient-specific CFD modelling into thoracoabdominal aortic aneurysm repair planning represents a step toward functional risk assessment that complements existing anatomical approaches. While current clinical decision-making appropriately relies on anatomical criteria—such as the number and location of segmental arteries to be sacrificed—this approach does not capture the haemodynamic interdependencies that govern actual spinal cord perfusion. Our CFD modelling approach builds upon anatomical assessment by quantifying how vessel exclusion translates into functional perfusion changes, considering factors such as geometric resistance and flow redistribution that are not apparent from anatomical imaging alone. This represents an evolution rather than a replacement of current methods, providing surgeons with additional quantitative data to complement clinical judgment in intervention planning and patient counselling.

 The observed 51.86\% reduction in spinal cord blood flow, combined with regional variations in wall shear stress metrics, illustrates the potential clinical relevance of this modelling approach as a foundation for future risk assessment tools. While validation studies are needed before clinical implementation, these findings suggest several areas where haemodynamic modelling might eventually inform clinical decisions. For instance, patients with pre-existing atherosclerotic burden in collateral pathways may represent a higher-risk population when segmental artery exclusion forces compensatory flow through potentially compromised vessels. Similarly, the differential haemodynamic responses observed across vascular territories could, following appropriate validation, contribute to decisions regarding staged procedures, adjunctive revascularisation strategies, or intervention timing based on individual patient physiology. As this modelling platform evolves to incorporate collateral circulation and adaptive mechanisms, it may provide increasingly sophisticated insights to complement traditional anatomical risk assessment methods.

Comparable CFD approaches have been successfully implemented in cardiovascular care, such as Heartflow (Heartflow, USA) for coronary artery assessment \cite{min2015noninvasive}. Our approach, based on standard imaging and open-source algorithms, aims to deliver similar personalised haemodynamic insights for thoracoabdominal aortic repair planning. The key challenge remains the lack of experimental validation metrics comparable to, for example, invasive fractional flow reserve in coronary applications.

\subsection{Limitations and future directions}  

A key limitation relates to the assumption taken on outflow boundary conditions. Uniform resistance values were assigned to all spinal cord-supplying arteries due to the lack of available patient-specific data. In practice, each artery supplies a distinct vascular bed with potentially unique resistance and compliance characteristics, and these differences are expected to influence both absolute and relative flow distributions. 

In addition to this simplification, the model does not account for the effects of physiological autoregulation, which modulates vascular tone in response to perfusion pressure changes \cite{evaniew2024interventions}. This may significantly alter flow redistribution, particularly following the exclusion of multiple spinal cord-supplying branches \cite{griepp2012anatomy}.

It is worth noting that the timepoint for the post-operative CTA (from which the post-operative model was created) in this case was one month. It could be argued that post-operative CTAs at different time points (e.g. one day or one year) would show a different picture with respect to the vascular adaptations which have occurred following the operation. This could influence interpretation regarding SCI probability. At present, little is known about vascular adaptive processes in this area, making it difficult to predict such influences. It is important to note that the ultimate aim of the pipeline is to support pre-operative prediction of SCI. In practice this means that no post-operative scan would be available at the time of prediction. Instead, expected anatomical changes following a planned operation would be inferred from the surgical plan, specifically from the vessels anticipated to be sacrificed, as determined during routine pre-operative planning.

Another important simplification is the exclusion of interactions with the cerebrospinal fluid (CSF) and venous compartments. Spinal perfusion pressure is influenced by the pressure gradients between arterial inflow, venous outflow, and CSF \cite{daouk2017heart}. Although such effects fall outside the scope of this initial model, they are physiologically relevant and should be included in future iterations to enhance realism.

The model also assumes rigid vessel walls, since we do not have data to parameterise the mechanical properties of the vessel wall. This may limit the accuracy of predictions in anatomical regions where wall compliance has a significant effect on local haemodynamics. Incorporating deformable wall properties through fluid–structure interaction modelling could refine the prediction of flow patterns, particularly under pulsatile conditions.

Flow estimates in this study were based on input values derived from the literature rather than direct patient-specific measurements. Although this approach provides a reasonable approximation, it reduces the precision of individual predictions. Ongoing efforts to acquire invasive haemodynamic data from segmental arteries in real patients will support more accurate boundary condition assignment and personalised simulation outputs.

Finally, this study presents results from a single patient and without a larger sample, including patients who did develop SCI, the model’s predictive value is speculative. While it demonstrates the feasibility and clinical potential of the modelling pipeline, broader validation across a larger cohort will be essential to confirm reproducibility, assess inter-patient variability, and determine predictive thresholds for spinal cord ischemia.

\section{Conclusion}

This study has provided the first means of objective quantification of haemodynamic properties through vessels supplying the spinal cord following aortic surgery using open-source software and standard clinical data. Modelling demonstrated approximately 50\% reduction in flow to the spinal cord after surgery. Key surface metrics of TAWSS, OSI, RRT and ECAP were reported and, owing to the holistic view of the aorta created by the pipeline, unexpected interregional differences were observed, most notably increased wall shear stress in visceral arteries compared to similar calibre vessels elsewhere. With further refinement and validation, the present tool has the potential to evolve into a non-invasive, pre-operative risk stratification method for spinal cord ischaemia, aligning with contemporary advances in personalised, computationally guided medicine.

 \bibliographystyle{elsarticle-num} 
 \bibliography{refs}

@article{rasiah2024optimizing,
  title={Optimizing medical management and risk factors},
  author={Rasiah, Michael Greshan and Modarai, Bijan},
  journal={Surgery (Oxford)},
  year={2024},
  publisher={Elsevier}
}

@article{nana2025editor,
  title={Editor's Choice--Role of antiplatelet therapy in patients managed for complex aortic aneurysms using fenestrated or branched endovascular repair},
  author={Nana, Petroula and Spanos, Konstantinos and Tsilimparis, Nikolaos and Haulon, St{\'e}phan and Sobocinski, Jonathan and Gallitto, Enrico and Dias, Nuno and Eilenberg, Wolf and Wanhainen, Anders and Mani, Kevin and others},
  journal={European journal of vascular and endovascular surgery},
  volume={69},
  number={2},
  pages={272--281},
  year={2025},
  publisher={Elsevier}
}

@article{rasiah2021medical,
  title={Medical management of risk factors for vascular disease},
  author={Rasiah, Michael Greshan and Modarai, Bijan},
  journal={Surgery (Oxford)},
  volume={39},
  number={5},
  pages={248--256},
  year={2021},
  publisher={Elsevier}
}

@article{rasiah2024need,
  title={Need for and update on clinical trials for uncomplicated type B aortic dissection},
  author={Rasiah, Michael Greshan and Abdelhalim, Mohamed Ahmed and Modarai, Bijan},
  journal={JVS-Vascular Insights},
  volume={2},
  pages={100130},
  year={2024},
  publisher={Elsevier}
}

@article{howard2015population,
  title={Population-based study of incidence of acute abdominal aortic aneurysms with projected impact of screening strategy},
  author={Howard, Dominic PJ and Banerjee, Amitava and Fairhead, Jack F and Handa, Ashok and Silver, Louise E and Rothwell, Peter M and Oxford Vascular Study},
  journal={Journal of the American Heart Association},
  volume={4},
  number={8},
  pages={e001926},
  year={2015}
}

@article{castro2019disparities,
  title={Disparities in contemporary treatment rates of abdominal aortic aneurysms across western countries},
  author={Castro-Ferreira, Ricardo and Lachat, Mario and Schneider, Peter A and Freitas, Alberto and Leite-Moreira, Adelino and Sampaio, S{\'e}rgio M},
  journal={European Journal of Vascular and Endovascular Surgery},
  volume={58},
  number={2},
  pages={200--205},
  year={2019},
  publisher={Elsevier}
}

@article{verzini2014current,
  title={Current results of total endovascular repair of thoracoabdominal aortic aneurysms.},
  author={Verzini, Fabio and Loschi, D and De Rango, P and Ferrer, C and Simonte, Gioele and Coscarella, C and Pogany, G and Cao, Piergiorgio},
  journal={The Journal of Cardiovascular Surgery},
  volume={55},
  number={1},
  pages={9--19},
  year={2014}
}

@article{abdelhalim2022multicenter,
  title={Multicenter Trans-Atlantic experience with fenestrated-branched endovascular repair of chronic postdissection thoracoabdominal aortic aneurysms},
  author={Abdelhalim, Mohamed A and Tenorio, Emanuel R and Oderich, Gustavo S and Modarai, Bijan},
  journal={Journal of Vascular Surgery},
  volume={75},
  number={6},
  pages={e123--e124},
  year={2022},
  publisher={Elsevier}
}

@article{dias2015short,
  title={Short-term outcome of spinal cord ischemia after endovascular repair of thoracoabdominal aortic aneurysms},
  author={Dias, NV and Sonesson, Bj{\"o}rn and Kristmundsson, Thorarinn and Holm, Hannes and Resch, Tim},
  journal={European Journal of Vascular and Endovascular Surgery},
  volume={49},
  number={4},
  pages={403--409},
  year={2015},
  publisher={Elsevier}
}

@article{SpinalCordInjuryStatisticalCenter2021,
   author = {National Spinal Cord Injury Statistical Center and Model Systems Knowledge Translation Center},
   title = {Spinal Cord Injury Facts and Figures at a Glance},
   url = {www.msktc.org/sci/model-system-centers.},
   year = {2021}
}

@article{min2015noninvasive,
  title={Noninvasive fractional flow reserve derived from coronary CT angiography: clinical data and scientific principles},
  author={Min, James K and Taylor, Charles A and Achenbach, Stephan and Koo, Bon Kwon and Leipsic, Jonathon and N{\o}rgaard, Bjarne L and Pijls, Nico J and De Bruyne, Bernard},
  journal={Cardiovascular Imaging},
  volume={8},
  number={10},
  pages={1209--1222},
  year={2015},
  publisher={American College of Cardiology Foundation Washington DC}
}

@article{updegrove2017simvascular,
  title={SimVascular: an open source pipeline for cardiovascular simulation},
  author={Updegrove, Adam and Wilson, Nathan M and Merkow, Jameson and Lan, Hongzhi and Marsden, Alison L and Shadden, Shawn C},
  journal={Annals of biomedical engineering},
  volume={45},
  pages={525--541},
  year={2017},
  publisher={Springer}
}

@incollection{wilson2018using,
  title={Using a science gateway to deliver simvascular software as a service for classroom instruction},
  author={Wilson, Nathan M and Marru, Suresh and Abeysinghe, Eroma and Christie, Marcus A and Maher, Gabriel D and Updegrove, Adam R and Pierce, Marlon and Marsden, Alison L},
  booktitle={Proceedings of the Practice and Experience on Advanced Research Computing: Seamless Creativity},
  pages={1--4},
  year={2018}
}

@article{ahrens200536,
  title={36-paraview: An end-user tool for large-data visualization},
  author={Ahrens, James and Geveci, Berk and Law, Charles and Hansen, C and Johnson, C and others},
  journal={The visualization handbook},
  volume={717},
  pages={50038--1},
  year={2005},
  publisher={Citeseer}
}

@book{ayachit2015paraview,
  title={The paraview guide: a parallel visualization application},
  author={Ayachit, Utkarsh},
  year={2015},
  publisher={Kitware, Inc.}
}

@article{fedorov20123d,
  title={3D Slicer as an image computing platform for the Quantitative Imaging Network},
  author={Fedorov, Andriy and Beichel, Reinhard and Kalpathy-Cramer, Jayashree and Finet, Julien and Fillion-Robin, Jean-Christophe and Pujol, Sonia and Bauer, Christian and Jennings, Dominique and Fennessy, Fiona and Sonka, Milan and others},
  journal={Magnetic resonance imaging},
  volume={30},
  number={9},
  pages={1323--1341},
  year={2012},
  publisher={Elsevier}
}

@incollection{stillaert2023lumbar,
  title={The Lumbar Artery Perforator Flap: A True Alternative in Autologous Breast Reconstruction},
  author={Stillaert, Filip BJL and Blondeel, Phillip and Van Landuyt, Koenraad},
  booktitle={Core Techniques in Flap Reconstructive Microsurgery: A Stepwise Guide},
  pages={231--242},
  year={2023},
  publisher={Springer}
}

@article{zhao2023aortic,
  title={Aortic flow is associated with aging and exercise capacity},
  author={Zhao, Xiaodan and Garg, Pankaj and Assadi, Hosamadin and Tan, Ru-San and Chai, Ping and Yeo, Tee Joo and Matthews, Gareth and Mehmood, Zia and Leng, Shuang and Bryant, Jennifer Ann and others},
  journal={European Heart Journal Open},
  volume={3},
  number={4},
  pages={oead079},
  year={2023},
  publisher={Oxford University Press US}
}

@article{westerhof1969analog,
  title={Analog studies of the human systemic arterial tree},
  author={Westerhof, Nicolaas and Bosman, Frederik and De Vries, Cornelis J and Noordergraaf, Abraham},
  journal={Journal of biomechanics},
  volume={2},
  number={2},
  pages={121--143},
  year={1969},
  publisher={Elsevier}
}

@article{xing2017distribution,
  title={Distribution of cardiac output to the brain across the adult lifespan},
  author={Xing, Chang-Yang and Tarumi, Takashi and Liu, Jie and Zhang, Yinan and Turner, Marcel and Riley, Jonathan and Tinajero, Cynthia Duron and Yuan, Li-Jun and Zhang, Rong},
  journal={Journal of Cerebral Blood Flow \& Metabolism},
  volume={37},
  number={8},
  pages={2848--2856},
  year={2017},
  publisher={SAGE Publications Sage UK: London, England}
}

@article{benim2011simulation,
  title={Simulation of blood flow in human aorta with emphasis on outlet boundary conditions},
  author={Benim, AC and Nahavandi, A and Assmann, A and Schubert, D and Feindt, P and Suh, SH},
  journal={Applied Mathematical Modelling},
  volume={35},
  number={7},
  pages={3175--3188},
  year={2011},
  publisher={Elsevier}
}

@article{lantz1981regional,
  title={Regional distribution of cardiac output: normal values in man determined by video dilution technique},
  author={Lantz, BM and Foerster, JM and Link, DP and Holcroft, JW},
  journal={American Journal of Roentgenology},
  volume={137},
  number={5},
  pages={903--907},
  year={1981},
  publisher={American Roentgen Ray Society}
}

@article{shiozawa2022blood,
  title={Blood pressure and coeliac artery blood flow responses during increased inspiratory muscle work in healthy males},
  author={Shiozawa, Kana and Kashima, Hideaki and Mizuno, Sahiro and Ishida, Koji and Katayama, Keisho},
  journal={Experimental Physiology},
  volume={107},
  number={9},
  pages={1094--1104},
  year={2022},
  publisher={Wiley Online Library}
}

@article{blanco2014blood,
  title={Blood flow distribution in an anatomically detailed arterial network model: criteria and algorithms},
  author={Blanco, Pablo J and Watanabe, Sansuke M and Dari, Enzo A and Passos, Marco Aur{\'e}lio RF and Feij{\'o}o, Ra{\'u}l A},
  journal={Biomechanics and modeling in mechanobiology},
  volume={13},
  pages={1303--1330},
  year={2014},
  publisher={Springer}
}

@article{itzchak1975external,
  title={External iliac artery blood flow in patients with arteriosclerosis obliterans},
  author={Itzchak, Yacov and Modan, Michaela and Adar, Rafael and Deutsch, Victor},
  journal={American Journal of Roentgenology},
  volume={125},
  number={2},
  pages={437--441},
  year={1975},
  publisher={American Roentgen Ray Society}
}

@article{bjorck2006blood,
  title={Blood-flow of the inferior mesenteric and internal iliac arteries among patients undergoing open surgery for abdominal aortic aneurysm},
  author={Bj{\"o}rck and Bergqvist},
  journal={Vasa},
  volume={35},
  number={1},
  pages={11--14},
  year={2006},
  publisher={Verlag Hans Huber}
}

@article{tran2024patient,
  title={Patient-specific computational flow simulation reveals significant differences in paravisceral aortic hemodynamics between fenestrated and branched endovascular aneurysm repair},
  author={Tran, Kenneth and Deslarzes-Dubuis, Celine and DeGlise, Sebastien and Kaladji, Adrien and Yang, Weiguang and Marsden, Alison L and Lee, Jason T},
  journal={JVS-Vascular Science},
  volume={5},
  pages={100183},
  year={2024},
  publisher={Elsevier}
}

@article{tsilimparis2017technical,
  title={Technical aspects of implanting the t-Branch off-the-shelf multibranched stent-graft for thoracoabdominal aneurysms},
  author={Tsilimparis, Nikolaos and Fiorucci, Beatrice and Debus, Eike Sebastian and Rohlffs, Fiona and K{\"o}lbel, Tilo},
  journal={Journal of Endovascular Therapy},
  volume={24},
  number={3},
  pages={397--404},
  year={2017},
  publisher={SAGE Publications Sage CA: Los Angeles, CA}
}

@article{de2019use,
  title={Use of bilateral Cook Zenith iliac branch devices to preserve internal iliac artery flow during endovascular aneurysm repair},
  author={de Marino, Pablo Marques and Botos, Balazs and Kouvelos, George and Verhoeven, Eric LG and Katsargyris, Athanasios},
  journal={European Journal of Vascular and Endovascular Surgery},
  volume={57},
  number={2},
  pages={213--219},
  year={2019},
  publisher={Elsevier}
}

@article{chung2022prevention,
  title={Prevention and management of spinal cord ischemia following aortic surgery: a survey of contemporary practice},
  author={Chung, Jennifer C and Lodewyks, Carly L and Forbes, Thomas L and Chu, Michael WA and Peterson, Mark D and Arora, Rakesh C and Ouzounian, Maral and Collaborative, Canadian Thoracic Aortic and Society, Canadian Cardiovascular Critical Care CANCARE},
  journal={The Journal of Thoracic and Cardiovascular Surgery},
  volume={163},
  number={1},
  pages={16--23},
  year={2022},
  publisher={Elsevier}
}

@article{griepp2012anatomy,
  title={The anatomy of the spinal cord collateral circulation},
  author={Griepp, Eva B and Di Luozzo, Gabriele and Schray, Deborah and Stefanovic, Angelina and Geisb{\"u}sch, Sarah and Griepp, Randall B},
  journal={Annals of cardiothoracic surgery},
  volume={1},
  number={3},
  pages={350},
  year={2012}
}

@article{daouk2017heart,
  title={Heart rate and respiration influence on macroscopic blood and CSF flows},
  author={Daouk, Jo{\"e}l and Bouzerar, Roger and Baledent, Olivier},
  journal={Acta radiologica},
  volume={58},
  number={8},
  pages={977--982},
  year={2017},
  publisher={SAGE Publications Sage UK: London, England}
}

@article{evaniew2024interventions,
  title={Interventions to optimize spinal cord perfusion in patients with acute traumatic spinal cord injury: an updated systematic review},
  author={Evaniew, Nathan and Davies, Benjamin and Farahbakhsh, Farzin and Fehlings, Michael G and Ganau, Mario and Graves, Daniel and Guest, James D and Korupolu, Radha and Martin, Allan R and McKenna, Stephen L and others},
  journal={Global spine journal},
  volume={14},
  number={3\_suppl},
  pages={58S--79S},
  year={2024},
  publisher={SAGE Publications Sage CA: Los Angeles, CA}
}

@article{lv2022artificial,
  title={An artificial intelligence-based platform for automatically estimating time-averaged wall shear stress in the ascending aorta},
  author={Lv, Lei and Li, Haotian and Wu, Zonglv and Zeng, Weike and Hua, Ping and Yang, Songran},
  journal={European Heart Journal-Digital Health},
  volume={3},
  number={4},
  pages={525--534},
  year={2022},
  publisher={Oxford University Press US}
}

@article{aucoin2023predictors,
  title={Predictors and outcomes of spinal cord injury following complex branched/fenestrated endovascular aortic repair in the US Aortic Research Consortium},
  author={Aucoin, Victoria J and Motyl, Claire M and Novak, Zdenek and Eagleton, Matthew J and Farber, Mark A and Gasper, Warren and Oderich, Gustavo S and Mendes, Bernardo and Schanzer, Andres and Tenorio, Emanuel and others},
  journal={Journal of vascular surgery},
  volume={77},
  number={6},
  pages={1578--1587},
  year={2023},
  publisher={Elsevier}
}

@article{carpenter2023nonlinear,
  title={On the nonlinear relationship between wall shear stress topology and multi-directionality in coronary atherosclerosis},
  author={Carpenter, Harry J and Ghayesh, Mergen H and Zander, Anthony C and Psaltis, Peter J},
  journal={Computer Methods and Programs in Biomedicine},
  volume={231},
  pages={107418},
  year={2023},
  publisher={Elsevier}
}

@article{kauppila2009atherosclerosis,
  title={Atherosclerosis and disc degeneration/low-back pain--a systematic review},
  author={Kauppila, LI},
  journal={European journal of vascular and endovascular surgery},
  volume={37},
  number={6},
  pages={661--670},
  year={2009},
  publisher={Elsevier}
}

@book{nichols2022mcdonald,
  title={McDonald’s blood flow in arteries: theoretical, experimental and clinical principles},
  author={Nichols, Wilmer W and O'Rourke, Michael and Edelman, Elazer R and Vlachopoulos, Charalambos},
  year={2022},
  publisher={CRC press}
}

@article{sotelo20163d,
  title={3D quantification of wall shear stress and oscillatory shear index using a finite-element method in 3D CINE PC-MRI data of the thoracic aorta},
  author={Sotelo, Julio and Urbina, Jesus and Valverde, Israel and Tejos, Cristian and Irarr{\'a}zaval, Pablo and Andia, Marcelo E and Uribe, Sergio and Hurtado, Daniel E},
  journal={IEEE transactions on medical imaging},
  volume={35},
  number={6},
  pages={1475--1487},
  year={2016},
  publisher={IEEE}
}

@article{riccardello2018influence,
  title={Influence of relative residence time on side-wall aneurysm inception},
  author={Riccardello Jr, Gerald J and Shastri, Darshan N and Changa, Abhinav R and Thomas, Kiran G and Roman, Max and Prestigiacomo, Charles J and Gandhi, Chirag D},
  journal={Neurosurgery},
  volume={83},
  number={3},
  pages={574--581},
  year={2018},
  publisher={LWW}
}

@article{mutlu2023does,
  title={How does hemodynamics affect rupture tissue mechanics in abdominal aortic aneurysm: Focus on wall shear stress derived parameters, time-averaged wall shear stress, oscillatory shear index, endothelial cell activation potential, and relative residence time},
  author={Mutlu, Onur and Salman, Huseyin Enes and Al-Thani, Hassan and El-Menyar, Ayman and Qidwai, Uvais Ahmed and Yalcin, Huseyin Cagatay},
  journal={Computers in biology and medicine},
  volume={154},
  pages={106609},
  year={2023},
  publisher={Elsevier}
}

@article{beckworth2018atherosclerotic,
  title={Atherosclerotic disease and its relationship to lumbar degenerative disk disease, facet arthritis, and stenosis with computed tomography angiography},
  author={Beckworth, William J and Holbrook, John F and Foster, Lisa G and Ward, Laura A and Welle, James R},
  journal={PM\&R},
  volume={10},
  number={4},
  pages={331--337},
  year={2018},
  publisher={Elsevier}
}

@article{xiao2014systematic,
  title={A systematic comparison between 1-D and 3-D hemodynamics in compliant arterial models},
  author={Xiao, Nan and Alastruey, Jordi and Alberto Figueroa, C},
  journal={International journal for numerical methods in biomedical engineering},
  volume={30},
  number={2},
  pages={204--231},
  year={2014},
  publisher={Wiley Online Library}
}

%% else use the following coding to input the bibitems directly in the
%% TeX file.

%% Refer following link for more details about bibliography and citations.
%% https://en.wikibooks.org/wiki/LaTeX/Bibliography_Management

% \begin{thebibliography}{00}

% %% For numbered reference style
% %% \bibitem{label}
% %% Text of bibliographic item

% \bibitem{lamport94}
%   Leslie Lamport,
%   \textit{\LaTeX: a document preparation system},
%   Addison Wesley, Massachusetts,
%   2nd edition,
%   1994.

% \end{thebibliography}

\appendix

\includegraphics[width=14cm]{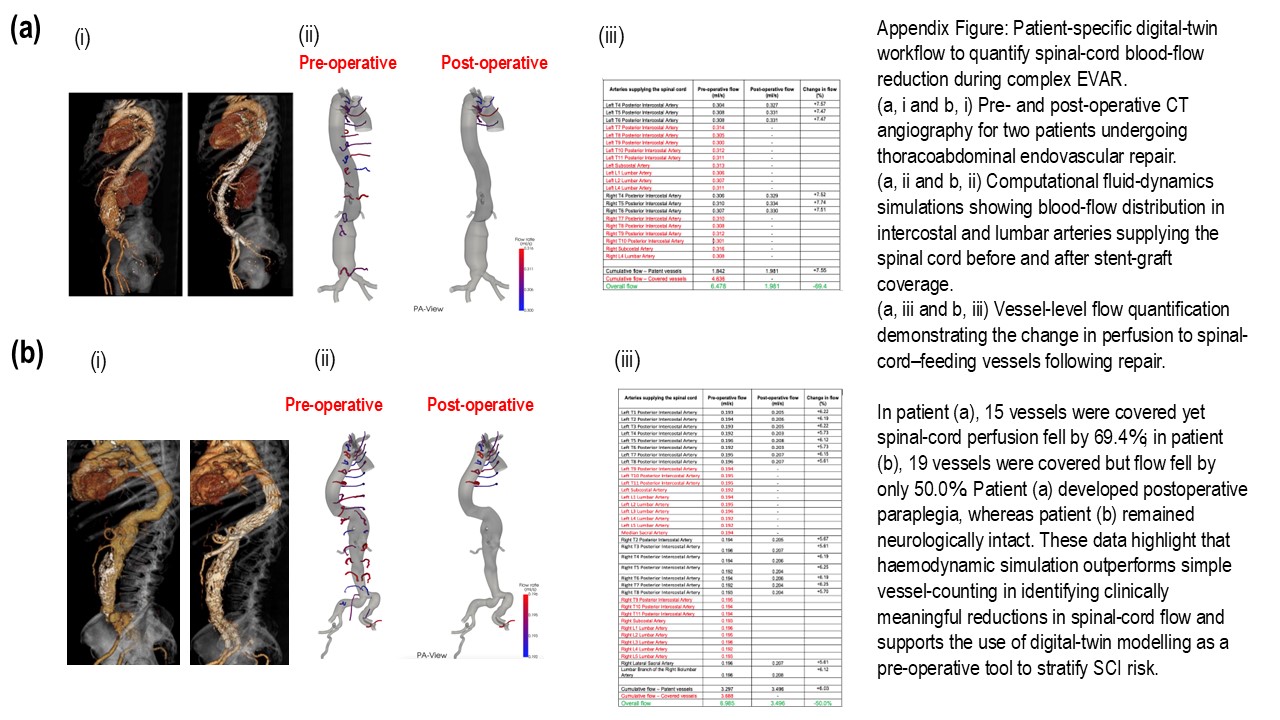}
\end{document}